\documentstyle[12pt]{article}
\begin{document}
{\large {\bf A Superspace Formulation for the Master Equation}}\\
\vspace{1cm}

{\large Everton M. C. Abreu $^{1,2}$ and   Nelson R. F. Braga $^{1}$} \\
\vspace{1cm}

$^1$ Instituto de F\'\i sica, Universidade Federal  do Rio de Janeiro,\\
Caixa Postal 68528, 21945  Rio de Janeiro,
RJ, Brazil\\

$^2$ Instituto de F\'\i sica, Universidade do Estado do Rio de
Janeiro,\\ Rua Sao Francisco Xavier 524,CEP 20550 Rio de Janeiro, RJ,
Brazil
\vspace{0.5cm}
\abstract
It is shown that the  quantum master equation of the
Field Antifield quantization method at one loop order can be translated into
the requirement of a superfield structure for the action.
The Pauli Villars regularization is
implemented in this BRST superspace and the case of anomalous gauge 
theories is investigated.
The quantum action, including Wess Zumino terms, shows up as one of
the components of a superfield that includes the  BRST anomalies in 
the other
component.  The example of W2 quantum gravity is also discussed.

\vfill\eject
\section{Introduction}

The gauge invariance principle is one of the basic ingredients
in the search for a description of the fundamental processes
involving elementary particles.
Gauge invariance is translated at the quantum level into the fermionic 
rigid BRST invariance\cite{BRST} and 
is important in  the proof of unitarity and renormalizability of 
field theories \cite{ZJ}.

The path integral quantization of gauge field theories poses some
interesting problems. The naive integration over all the field configurations
would lead to an over counting of physically equivalent ones.
A mechanism of factoring out this
over counting, at least for some special kind of gauge field theories
was proposed by Faddeev and Poppov \cite{FP}.

The  Batalin Vilkovisky (BV) formalism\cite{BV,HT,GPS}, also called
field antifield quantization, is a Lagrangian BRST procedure that
generalizes the Faddeev Poppov mechanism  and also incorporates the
idea\cite{ZJ} of including sources of the BRST transformations
as independent variables as an important tool for deriving the Ward
identities.  The formalism is defined in  an extended space that
includes the fields and also the BRST sources, called antifields.  In
this space, the Ward identities, representing the BRST invariance of
the vacuum functional, can be cast into a general expression, called
master equation.  One of the main goals of this general approach is
that for the case of reducible gauge theories  it
furnishes a systematic way of building up the non trivial ghost for
ghost structure.  Also quantum corrections from the path integral
measure, for anomalous gauge theories, can be calculated as long as a
regularization procedure is introduced\cite{TPN}.

It is known that superspace formulations for gauge field theories can
be build up in such a way that the BRST transformations are realized
as translations in a Grasmannian coordinate\cite{SUP}. It has also
recently been shown \cite{BD} that the BV formalism  at classical
level (zero order in $\hbar$) can also be cast in such a BRST
superspace  form.

If we use the standard BV formulation,  when we go to the BRST
superspace we will in general find constrained superfields.  As will
be discussed in section {\bf 2 }, functional differentiation with respect
to superfields will be essential in finding a superspace version for
the operator $\Delta$, but   for constrained superfields we can not
find a general definition for functional derivatives.  In order to
overcome this obstacle we will consider an alternative derivation for
the BV action proposed in
\cite{COL}. In this so called collective approach to BV, the set of
fields of the classical theory and also the ghosts, antighosts and
auxiliary fields associated to the original gauge symmetries is
trivially doubled.  This leads to new (trivial) shift symmetries.
These extra symmetries can then be gauge fixed (adding new ghosts
antighosts and auxiliary fields) in such a way that the BV action is
recovered after the extra fields are integrated out. The antighosts
of the trivial symmetries are identified as playing  the role of the
associated antifields.  The transformations of the fields are chosen
in such a way that at least for the superfields  that will be
relevant in building the superspace $\Delta$ operator the components
will be independent.

In the present article we will investigate a superspace version of
the field antifield formalism at one loop order in $\hbar$.  We will
find out that the master equation implies a certain structure
for the superfield associated to the quantum action.
We will also see that the Pauli Villars regularization
can be formulated in this BRST superspace.  A well defined meaning
can thus be given to the superfield structure of the quantum action.
The example of W2 gravity will nicely illustrate the formulation.
 
The article is organized as follows: in section {\bf 2} we discuss
the superspace formulation for the BRST symmetry, explaining why do
we use the collective field approach rather then the usual BV.  In
section {\bf 3} we review the superspace formulation for BV at
classical level.  In section {\bf 4} we present the general form of
the master equation and of the superfield associated to the quantum
action at one loop order. Section {\bf 5} is devoted to the Pauli
Villars regularization in superspace.  In section {\bf 6} we show the
form of the superfield action at one loop order. The  example of W2
gravity is discussed in section {\bf 7} and section {\bf 8} contains
some concluding remarks.

\section{ BRST superspace}

Superspace formulations for the BRST transformation\cite{SUP} are
obtained by associating to each field $\phi (x)$ a (BRST) superfield
of  the form:

\begin{equation}
\label{SBRST}
 \Phi(x,\theta ) = \phi (x) + \theta \delta \phi (x)
\end{equation}

\noindent where $\delta\phi (x)$ is the BRST transformation of
$\phi(x)$. The BRST transformations are then realized as translations
in the $\theta$ variable.

\begin{equation}
\label{TRANS}
\delta \Phi(x,\theta ) ={\partial\over \partial \theta} 
\Phi(x,\theta )
\end{equation}

In order to apply this idea to the BV Master equation at one loop
order, it is crucial that we define in a very precise way functional
derivatives with respect to superfields.  Let us start considering a
general bosonic superfield of the form:

\begin{equation}
\Lambda (x , \theta ) = A (x) + \theta B (x)
\end{equation}

\noindent where $A(x)$ and $B(x)$ are independent
quantities:

\begin{equation}
\label{INDEP}
{\delta B(x)\over \delta A(x^{\prime})} \,=\,0\,\,\,\,;
\,\,\,\,\,\,
{\delta A(x)\over \delta B(x^{\prime})}\,=\,0
\end{equation}

If we define a functional derivative so as to satisfy:

\begin{equation}
\label{SDEL}
{\delta \Lambda (x,\theta)\over
\delta \Lambda (x^\prime ,\theta^\prime )} = \delta (x - x^\prime )
\delta (\theta^\prime - \theta ) = \big( \delta (x - x^\prime )
\big) \,\,(\theta^\prime - \theta )
\end{equation}

\noindent we recover, for superspace functionals,  the usual
interpretation of the functional derivative, as for example:

\begin{equation}
\label{SQF}
{\delta \over \delta \Lambda (x ,\theta )}
\int dx^\prime \int d\theta^\prime 
\Big( \Lambda (x^\prime , \theta^\prime ) \Big)^2 
= 2 \Lambda ( x , \theta )
\end{equation}

\noindent it is important to remark that this superfield functional
derivative has a Grassmanian parity opposite to the associated field.
We can also express the derivative with respect to the superfield in
terms of derivatives with respect to it's components:

\begin{equation}
\label{COMP}
{\delta \over 
\delta \Lambda (x ,\theta )} \, = 
\,{\delta \over 
\delta B (x )} 
\,+ \, \theta {\delta \over 
\delta A (x )} 
\end{equation}

Considering now the BRST superfields of eq. (\ref{SBRST}),  
condition (\ref{INDEP}) of independent components clearly does
not hold in  general. Actually one loop level corrections
 are just
associated to the  contributions from (singular) terms like 

\begin{equation}
{\delta (\delta_{BRST}\phi)\over \delta\phi}
\end{equation}

\noindent we will thus be dealing in general with constrained superfields.
\bigskip

A simple way to realize that for the BRST superfields of
eq.(\ref{SBRST}) the naive application of superfield functional
derivatives would lead to contradictory results is to consider the
case of a field with vanishing BRST transformation, as for example
the ghost fields in QED. For the associated superfield,  with no
$\theta$ component, if we  try to define a functional derivative
satisfying (\ref{SDEL}) we would arrive at the contradiction that the
functional in (\ref{SQF}) vanishes and would have a non vanishing
derivative.

In standard supersymmetry a similar situation happens when one
considers chiral or antichiral superfields. A functional derivative
can be defined for these special constrained superfields but
(\ref{SDEL}) is replaced by an appropriate version that takes the
particular constraint into account. It is however impossible to
define functional derivatives (and also path integration) for general
constrained superfields\cite{GGRS}.

In order to have a general superspace version for the BV master
equation at one loop order  we should find a superspace version of
the operator $\Delta$, that involves two functional derivatives. As
we have seen, just substituting fields by BRST superfields would be
meaningless unless one finds a  general definition for their
functional derivatives.  An interesting way to overcome this
obstruction is to use the so called collective field approach to BV.
As we will see in the following sections, the BRST algebra in this
case will be such that, at least for the  fields that will be used in
the $\Delta $ operator, conditions (\ref{INDEP}) hold.

\section{Superspace Formulation at Classical Level}

In reference \cite{BD} a superspace formulation for the
collective field approach to the
Batalin Vilkovisky action at the order zero in $\hbar$ 
was presented for the case of the Yang Mills theory.
Here we will briefly review this formulation, presenting it in a
general way for gauge theories with closed gauge algebra.

Considering a gauge field theory characterized by a classical action
$S_0[\phi^i] $ we introduce ghosts, antighosts and auxiliary fields
associated to the original gauge invariance of  $S_0$ in the usual
way. The new enlarged set of fields is then denoted as $\phi^A$.
These fields realize the BRST algebra represented as:

\begin{equation}
\label{OGS}
\delta_0 \phi^A = R^A \,[\phi ]
\end{equation}

Then we introduce a new set of fields called collective fields
$\tilde \phi^A$ and replace everywhere $\phi^A$ by $\phi^A - \tilde
\phi^A$.  This way we double the field content of the theory and at
the same time associate to each field a new trivial shift symmetry.
In order to gauge fix these new symmetries we introduce new ghosts,
antighosts and auxiliary fields, represented respectively as:
$\pi^A$, $\phi^{\ast\,A}$ and $B^A$.  We have a large freedom in
choosing the BRST transformations for this enlarged set of fields.
Following \cite{COL} let us define the enlarged BRST algebra as :

\begin{eqnarray}
\label {ALG}
\delta \phi^A &=& \pi^A\nonumber\\
\delta \tilde \phi^A &=& \pi^A - R^A [\phi -\tilde \phi ]
\nonumber\\
\delta \pi^A &=& 0\nonumber\\
\delta \phi^{\ast\,A} &=& B^A\nonumber\\
\delta B^A &=& 0
\end{eqnarray}

\noindent and the total action as:

\begin{equation}
\label{A1}
S_{col.} = S_0 [\phi^i -\tilde \phi^i ] - \delta (\phi^{\ast\,A}
\tilde \phi^A ) + \delta \psi [\phi^A]
\end{equation}

\noindent where $\psi [\phi^A]$ is a fermionic functional
representing the gauge fixing of the original symmetries (\ref{OGS}).
The BV gauge fixed classical action is obtained if one functionally
integrates the vacuum functional associated with $S$ over $\pi^A$,
$\tilde
\phi^A$ and $B^A$.

Now, the BRST superspace formulation is obtained introducing the
superfields

\begin{eqnarray}
\label{EXTALG}
\Phi^A (x,\theta ) &=& \phi^A (x) + \theta \pi^A (x) \nonumber\\
\tilde \Phi^A (x,\theta ) &=& \tilde \phi^A (x) + \theta 
( \pi^A (x) - R^A [\,\phi -\tilde \phi \,]\, )\nonumber\\
\Phi^{\ast\,A} (x,\theta ) &=& \phi^{\ast\,A} (x) + \theta B^A (x) \nonumber\\
\end{eqnarray}

\noindent We can also associate superfields to the ghosts and the
auxiliary fields of the shift symmetry but they would have a trivial
structure: $\underline \Pi^A (x,\theta )\, =\,  \pi^A (x) \,\,\,$,
$\,\, \underline B^A (x,\theta) \,=\, B^A (x)\,$.

Considering the set (\ref{EXTALG}) we can define a
superfield action as:

\begin{equation}
\label{ST}
{\underline S}_{col.} = \, \,\, S_0 [\Phi^i -\tilde {\Phi}^i ]\,
\,-\, {\partial\over \partial \theta} \,\{
\Phi^{\ast\,A}
\tilde {\Phi}^A \, + \,  \psi [\Phi^a] \,\,\}
\end{equation}

This object actually has a trivial superspace structure as it's
$\theta$ component is zero ($\,\,{\underline S}_{col.} =
S_{col.}\,\,$).  It may thus seem meaningless at this stage to
associate a superfield to the action.  We will see however in the
next section that when higher order terms in $\hbar$ are taken into
account the situation is rather different.  At Classical level,
$S_{col.}$ is BRST invariant, therefore the associated superfield
must have a zero $\theta$ component, expressing what we will see in
the next section to correspond to the zero order term of the master
equation.  We will see in the next section that at higher in $\hbar$
the quantum action is not BRST invariant and the associated
superfield structure will not be trivial.

Concluding this section we remark that in the collective field
approach of \cite{COL}, presented here, the fields $\phi^{\ast\, A}$
that play the role of antifields are  substituted by the gauge fixing
conditions,  after integration over the auxiliary fields. In other words, 
we get the BV gauge fixed action. One may
however be interested in an action that still involves the antifields
as for example if one wants to build up an effective action in terms
of classical fields and antifields
\cite{HT}. In order to show that the collective field approach can also 
reproduce this non completely gauge fixed result we can add to 
$\underline S_{col.}$ the term:

\begin{equation}
{\partial \over \partial \theta}\,\,\Big( 
\Phi^A \, \underline \varphi^A \,\Big)\,=\, \varphi^A \,\pi^A
\end{equation}

\noindent where $\varphi^A$ are BRST invariant external fields
with parity opposite to that of $\phi^A$.  Integration over the auxiliary 
fields would recover the BV action with external antifields as:

\begin{eqnarray}
 exp\{ {i\over \hbar } S_{BV} [ \phi^A \, , \, {
\partial \psi \over \partial \phi^A }
\, +\, \varphi^A ] \} = \,\,\,\,\,\,\,\,\,\,\,\,\,\,\,\,\,\,\nonumber\\
= \int D\tilde\phi^A  D\phi^\ast D \pi^A 
D B^A
 exp \{ {\underline S}_{col.}\, +\, {\partial \over \partial \theta}
\,\Phi^A \underline \varphi^A \,\,\}
\end{eqnarray}

\noindent for $\varphi^A = 0$ this reduces to the gauge fixed BV action.
For $\psi = 0$ we get the non gauge fixed action.

  \section{Superspace version of the Master equation}

We will first investigate the BRST variation of the quantum action in
the standard field antifield quantization method. Then we will see
the corresponding behavior in the collective field approach.  We
will consider the case of gauge theories with closed gauge algebra.
Anomalies may in general have a non trivial dependence on the
antifields \cite{AnAnt1,AnAnt2}. We will however consider here a
regularization procedure that will only provide for the antifield
independent part of $\Delta S$. For gauge theories with closed
algebras one can consistently consider this part separately\cite{GP}.
Recently proposed non-local\cite{Paris} or  antifield
dependent\cite{GKPZ} regularization procedures are also out the scope
of our present superspace formulation.

The condition of gauge independence of the vacuum functional

$$ Z_\Psi = \int\prod D\Phi^A exp\left( {i\over\hbar} W(\phi^A,
\phi^{*A} = {\partial\psi\over \partial \phi^A}\right) $$

\noindent is translated into the so called (quantum) master equation:

\begin{equation}
\label{Master}
{1\over 2}(W,W) = i\hbar\Delta W
\end{equation}

\noindent  where the antibracket is defined as:
 $(X,Y) = {\partial_rX\over
\partial\phi^a} {\partial_lY\over\partial\phi^{\ast a}}
- {\partial_rX\over \partial\phi^{\ast a}}
  {\partial_lY\over \partial\phi^a}$
and the operator Delta as:  $\Delta \equiv
{\partial_r \over\partial \phi^a }{ \partial_l \over \partial \phi^\ast_a }\;$

The quantum action can be expanded in a power series in $\hbar$ as: $
W(\phi^A ,\phi^{*A} ) = S(\phi^A ,\phi^{*A} ) +
\sum_{p=1}^\infty \hbar^p M_p (\phi^A ,\phi^{*A} )$
we will be concerned here just with the first two terms, since we are
considering just one loop corrections.  In order to investigate the
behavior of $W$ with respect to BRST transformations, let us consider
the BRST transformation for some quantity X in the standard BV
language \cite{H}:

\begin{equation}
\label{transf}
\delta X = (X,W) -i\hbar \Delta X 
\end{equation}

\noindent if we choose $X = W $ and consider  that we are dealing with a 
non anomalous theory such that the master equation 
is satisfied 
we get from (\ref{transf}) and (\ref{Master}):

\begin{equation}
\label{TA} 
\delta W = i\hbar \Delta W 
\end{equation}

\noindent This condition is equivalent to the master equation. 

Under our present  assumption that $M_1$ does not depend on the
antifields we have (up to one loop order only)

\begin{eqnarray}
\delta S &=& 0 \nonumber\\
\delta M_1 &=& i \Delta S 
\end{eqnarray}

Going now to the collective field case. We see from  (\ref{ST}) that
the extended action $S_{col.}$ is also BRST invariant.  For the
action of the $\Delta$ operator we get a similar result in the
collective field approach and in the standard one:

\begin{equation}
\Delta S = \Delta S_{col.} = 
{\partial_r \over\partial \phi^A }
{ \partial_l \over \partial \phi^{\ast\,A} }\; S_{col.}
\end{equation}
 
\noindent it should be noted, however that in the collective
field case $\phi^{\ast\,A}$ are not antifields but rather antighosts of the shift symmetry. 
Therefore, at one loop order, we  must build up a superfield 

\begin{equation}
{\underline M}_1 [ \Phi^A - \tilde\Phi^A ]  = M_1 [ \phi^A - \tilde \phi^A ]
 + \theta i \Delta S_{col.} 
\end{equation}

\noindent and the general form of the superfield action will be:

\begin{equation}
\label{SA}
\underline W = W + \theta i\hbar \Delta W 
\end{equation}

Actually this expression for the superfield $\underline W $ is just
formal, in the same way as the master equation (\ref{Master}) itself.
We can only have a precise interpretation for terms involving the
operator  $\Delta $, where two functional derivatives act on the same
space time point if some regularization procedure is applied
\cite{TPN}.  We will show in the next section how the Pauli Villars
regularization procedure can be implemented in this superspace.

Let us now define, in the collective field space, the operator 

\begin{equation}
\label{DS}
\underline \Delta \equiv 
\int dx  \int d\theta \int d\theta^\prime\,{\delta_r \over 
\delta \Phi^A (x,\theta ) }\,\,
{\delta_l \over 
\delta \Phi^{\ast\,A} (x ,\theta^\prime \,)}
\end{equation}

\noindent where we have indicated explicitly the integrations over
space-time and Grassman variables, omitted in the previous
expressions, because of the non trivial form (the functional
derivatives are taken in the same space time point  but in different
Grassman coordinates).

Looking at  (\ref{EXTALG}) we see that the superfields involved in
$\underline \Delta$ satisfy   (\ref{INDEP}). Therefore the functional
derivatives are well defined and we can also use the decomposition in
components (\ref{COMP}) in order to calculate:

\begin{equation}
\label{SB}
\underline \Delta \,\,{\underline S}_{col.} = 
\underline \Delta
\int dx \,
\int d\theta\,\,\Big( \,-\,\Phi^{\ast\,A} (x,\theta )
\tilde {\Phi}^A (x,\theta ) \Big)
\,=\,  \int dx\,\, {\delta_r ( R^A (x) )
\over \delta  \phi^A (x) }
\end{equation}

\noindent that is precisely the result that one obtains in the 
standard BV formalism if the $\Delta $ operator is naively
applied to the Classical Action. We have thus found a superspace 
representation for this operator. 

The master equation in superspace then reads:

\begin{equation}
\label{SMaster}
{\partial \over \partial \theta}  \underline W \,=\, 
i\hbar \underline \Delta \underline W 
\end{equation}

\noindent or order by order:

\begin{equation}
{\partial \over \partial \theta}  \underline S \,=\, 
0 \,\,\,\,;\,\,\,\,
{\partial \over \partial \theta}  \underline M_1 \,=\, 
i \underline \Delta \underline S 
\end{equation}
 
At this point one could question about the lack of an antibraket
structure in the present superspace approach.  However, looking at
(\ref{TRANS}) and (\ref{SMaster}) one realizes that the role of
generator of BRST transformations is essentially played by the
differentiation with respect to $\theta$.  Therefore,  enlarging the
configuration space with the variable $\theta$, we are equipping it
with Grassmanian translations that reproduce the effect of the
antibrackets. So, this structure is not necessary and would be
redundant.

\section{One loop order regularization in superspace}

The Pauli Villars  regularization procedure is the most suitable for 
the BV formalism at one loop order\cite{GPS,TPN,TP,DJ}.
We will consider, for simplicity, the case of just one Pauli Villars (PV)
field associated to each field of the theory. In some cases  one 
needs a set of PV fields but this modification 
would not change the superspace structure, as it will be seen in the example.
In the present superspace formulation the field content of the theory
is  enlarged by the addition of the collective fields and the gauge
fixing structure of the associated shift symmetries.  We will build
up a Pauli Villars superfield action corresponding to a collective
field version of the standard PV action, or equivalently, to a PV
partner of action $\underline S_{col.}$ of eq. (\ref{ST}):

\begin{eqnarray}
\underline S_{PV} &=&   {1\over 2} (\underline \chi^A - 
\underline {\tilde \chi}^A ) (T{\sl O})_{AB}
(\underline \chi^B - \underline {\tilde \chi}^B )\nonumber\\
&-&{1\over 2} M (\underline \chi^A - 
\underline {\tilde \chi}^A ) T_{AB}
(\underline \chi^B - \underline {\tilde \chi}^B ) 
- {\partial\over \partial \theta}\,
(\,\underline \chi^{\ast\,A} \underline {\tilde \chi}^A )
\end{eqnarray}

\noindent as in
\cite{TPN}, the matrix $T$ is an arbitrary invertible one
while $T {\sl O}$ is :

\begin{equation}
(T{\sl O})_{AB} = {\partial_l \over \partial \Phi^A}
{\partial_l \over \partial \Phi^B} S^{\prime}
 (\Phi^A, {\partial \psi\over \partial \phi^A} )
\end{equation}

\noindent where $S^{\prime} (\Phi^A, {\partial \psi\over \partial \phi^A}\,)$
  is obtained from the original action (\ref{A1}) after removing the
collective fields

\begin{equation}
exp ( {i\over \hbar} S^{\prime} (\Phi^A, {\partial \psi\over \partial
\phi^A })) )  = \int D \tilde\phi^A\, D \pi^A \, D B^A exp\,\Big(
{i\over\hbar}\, \, {\underline S}_{col.} \Big)
\end{equation}

In order to build up the PV superfields we must define their enlarged
BRST algebra. We define the matrix

\begin{equation}
K_{AB} \,=\, {\partial_l \over \partial \Phi ^{\ast \,A}} 
{\partial_r \over \partial \Phi ^B } 
{\underline S}_{col.}
\end{equation}

\noindent where $\underline S$ is the action of eq. (\ref{ST}), that
actually has no $\theta$ component, and  impose that the non extended
(without collective fields) BRST algebra for the PV fields reads:

\begin{equation}
\delta_{_0} \chi^A\,=\, K_{AB} \chi^B
\end{equation}

Following the steps of section {\bf 2}  we find the enlarged algebra for 
the PV fields and build up the associated superfields:

\begin{eqnarray}
\label{PVSUP}
{\underline \chi}^A (x,\theta ) &=& \chi^A (x) + \theta \,
\,\pi^{[\,\chi\,]A\,}
\nonumber\\
{\tilde {\underline \chi}}^A (x,\theta ) &=& \tilde \chi^A (x) + \theta 
( \,
\,\pi^{[\,\chi\,]A\,} -  K_{AB} (\chi^B - \tilde\chi^B ) \, )\nonumber\\
{\underline \chi}^{\ast\,A} (x,\theta ) &=& \chi^{\ast\,A} (x) + 
\theta B^{\,[\,\chi\,]\,A}
\end{eqnarray}

As usual, the PV fields are defined formally in such a way that their 
one loop contributions have a minus sign relative to the original fields.
The action of the operator $\underline \Delta $  on the 
regularized  total action is thus:

\begin{eqnarray}
\label{DSR}
\underline \Delta (\underline S + \underline S_{PV} ) &\equiv&
\int dx \int d\theta \int d\theta^\prime \Big( {\delta_r \over 
\delta \Phi^A (x,\theta ) }
{\delta_l \over 
\delta \Phi^{\ast\,A} (x ,\theta^\prime )}\nonumber\\
&+& {\delta_r \over 
\delta \underline \chi^A (x,\theta ) }
{\delta_l \over 
\delta \underline \chi^{\ast\,A} (x ,\theta^\prime )}\Big)
\,\,(\,\underline S + {\underline S}_{PV}\, )\,\, =\,\, 0
\end{eqnarray}

The regularized form of $\Delta S$ in the non superspace case
 shows up in the violation of the zero order master equation 
associated  to the presence of the mass term.
In superspace this absence of
BRST invariance of the total (regularized) classical action $S_{T} =
S + S_{PV}$ is translated into the presence of a $\theta$ component in
the corresponding superfield:

\begin{equation}
\underline S_{T} = \underline S + \underline S_{PV} = 
S_T + \theta \delta S_T
\end{equation}

The general form of $\delta S_T$ is

\begin{equation}
\label{1}
\delta S_T = M \, \big( (\chi^A - \tilde\chi^A ) T_{AC} K^C_{\,\,B}
(\chi^B - \tilde\chi^B ) + {1\over 2 } (\chi^A - \tilde \chi^A ) 
\delta T_{AB} \,(\chi^B -\tilde \chi^B) \big)
\end{equation}

\noindent Integration over the fields $\pi^{[\,\chi\,]\,A\,}
\,,\,  B^{\,[\,\chi\,]\,A}$ and $\tilde \chi^A$ removes the 
extended collective field structure, recovering  the
usual result as in \cite{TPN}, that corresponds in (\ref{1})
 just to the absence 
of the collective tilde fields.
The next step would be to integrate over the PV fields. We will
not repeat this procedure here as it is widely discussed in the
literature\cite{GPS,TPN,GP,TP,DJ}. Let us assume that a regularized 
form of the BRST change in the total action 
$(\delta S_T\,)_{\,Reg.} $ was calculated. 
Using  eq. (\ref{TA}) up to one loop order 
terms
we find the relation between the BRST variation of the regularized action
and the desired regularized $\Delta S$:

\begin{equation}
 i\,\hbar\,\Big( \Delta S \Big)_{Reg} \,=\, \Big(\, \delta S_T \Big)_{Reg}
\end{equation}

\section{Anomalous gauge theories}
Genuine anomalies are characterized by a violation of the master
equation\cite{TPN}. For this kind of theories, it is not possible to find an 
$M_1$ term in the original space of fields and antifields, such that 
the master equation (\ref{Master}) is satisfied. 
As already explained, in the present superspace formulation we are 
considering the particular case in which the regularization procedure
provides just the antifield independent contributions to $\Delta S$.
We will thus consider only antifield independent anomalies and Wess
Zumino terms.  Under this assumption, the violation of the master
equation is of the form:

$$ \Delta W + {i\over 2\hbar} (W,W) = {\bf A} = c^\alpha A_\alpha $$

\noindent The symmetries associated to the ghosts $c^\alpha$ are said 
to be
broken by the anomalous behavior of the path integral measure.  In
this case the BRST transformation for the action has the form

$$ \delta W = i\hbar \Delta W - 2i\hbar c^\alpha A_\alpha $$

The superfield associated to the quantum action will then look like

\begin{equation}
\label{SAn}
\underline W = W + \theta ( i\hbar\Delta W - 2i\hbar c^\alpha A_\alpha )
\end{equation}

It is interesting to discuss in our  superspace formulation  the
mechanism of restoring gauge invariance by the inclusion of additional
degrees of freedom associated to the (broken) gauge group,
proposed by Faddeev and Shatashvili\cite{FS}. In the
BV  formalism this mechanism is implemented by enlarging the field
antifield space\cite{BM} by including fields associated to the gauge
group.  This way one can find a description for a (potentially)
anomalous gauge theory in which the classical symmetries are realized
at the quantum level, at the cost of some of the gauge group degrees
of freedom becoming dynamical. In the present one loop level superspace
formalism this, so called Wess Zumino mechanism, corresponds to finding 
out a superfield (involving the additional field antifield pairs):

\begin{equation}
{\underline M}_1  =  M_1 + \theta \Big( i \Delta S \Big)_{Reg}
\end{equation}

\noindent such that the superfield action takes the non anomalous
form (\ref{SA}) and one says that the anomalies have been canceled.

As remarked in \cite{JST},
another interesting interpretation for this mechanism of canceling the
anomalies in the original gauge symmetries, is that actually the
anomalies are not canceled but shifted to a trivial sector of
symmetries. One arrives at this result considering that the extra
fields that realize the Wess Zumino mechanism are not present at the
classical level and one should thus include in the classical action a
gauge fixing term associated to the invariance with respect to any
shift in these fields .  Taking this point of view, the  form of the
superfield action, even after the Wess Zumino term is included,
is still as in (\ref{SAn}) 
the only difference is that the anomalies are shifted  to extra ghosts  
$d^\beta$ associated to the (broken) trivial shift symmetries of the 
additional fields.
This situation will be clarified  in the example of the next section.

\section{Example}

Let us consider  W2 gravity theory as an interesting example of an
anomalous gauge theory that can be cast into the present superspace
BV formulation.
The classical theory is described by 

$$S_0=  {1\over 2\pi} \int d^2x \left[ \partial\phi 
\overline\partial\phi - h (\partial\phi )^2 \right]  $$

The BRST algebra associated to this theory is:

\begin{eqnarray}
 \delta_0 \phi &=& c \partial\phi \nonumber\\
\delta_0 h &=& \overline \partial c - h\partial c
 + \partial h c \nonumber\\
\delta_0 c &=& (\partial c ) c  
\end{eqnarray}

\noindent where $c$ is the ghost associated with the original gauge
invariance of $S_0$.

Now we follow the procedure of section {\bf 2} and enlarge the field
content of the theory introducing the collective fields associated to
$\phi\,,h\,$ and $c$, represented by tilde fields, the ghosts,
antighosts and auxiliary fields. Then we build up the superfields 
$ \Phi(x,\theta ) , \tilde\Phi(x,\theta ) , \Phi^\ast (x,\theta )$ ,
$H(x,\theta ) , \tilde H(x,\theta ) , 
H^\ast (x,\theta )$ , $\eta (x,\theta ) = c(x) + \theta \delta c (x) , 
\tilde\eta (x,\theta ) ,
\eta^\ast (x,\theta ) $ as in (\ref{EXTALG}).

We will adopt the notation $\,\sigma^\prime\, = 
\,\sigma - \, \tilde\sigma\,\,$ for all
fields and superfields in the rest of the section.
The  superfield action at classical level is:

\begin{equation}
\underline S = \underline S_0 + \underline S_1 + \underline S_2
\end{equation}

\noindent with the collective field version of the classical action:

$$\underline S_0 = {1\over 2\pi}  
\int d^2x \,\, \left[ \partial \Phi^\prime \,
\overline\partial \Phi^\prime - H^\prime 
(\partial \Phi^\prime )^2 \right] $$

\noindent the gauge fixing of the shift symmetry: 

$$\underline S_1 = - {\partial \over \partial\theta} \int d^2x \left[
\Phi^{\ast} \tilde \Phi + H^\ast \,\tilde H + \eta^\ast\, \tilde\eta
\right]$$

\noindent and the gauge fixing for the original symmetry:

$$\underline S_2 = {\partial\over \partial\theta} \int d^2x  \, \psi
(\Phi , H, \eta ) $$

To realize the Wess Zumino mechanism one includes an extra field
$\rho$ transforming according to the original gauge transformations 
associated to the ghost $c$ and  also with an additional shift symmetry
associated to an extra ghost $d\,$, representing the absence of this
field at classical level \cite{DJ}:

$$ \delta_{_0} \,\rho = \partial c + c\, \partial\rho + d $$

We introduce a collective field structure for this field, and
associate to it the superfields  $\Omega (x,\theta ) ,
\tilde \Omega (x,\theta ) , \Omega^{\ast} (x,\theta ) $ as in (\ref{EXTALG}).
Usually one is interested in calculating the contributions from the
matter fields $\phi$ only, considering the field $h$ as a background. 
Therefore we  introduce, as in \cite{JST}, a Pauli Villars field associated to
$\phi$ that will be represented as  $\chi$
and define  the PV superfield action:

\begin{equation}
\label{PVAction}
{\underline S}_{\,PV} = {\underline S}_{\,PV\,0} +
 {\underline S}_{\,PV\,1} + {\underline S}_M
\end{equation}

\noindent with the first two terms analogous to the corresponding
terms of the original fields:

\begin{equation}
{\underline S}_{\,PV\,0} = {1\over 2\pi}  
\int d^2x \,\, \Big( \partial \underline \chi^\prime
\overline\partial {\underline \chi}^\prime 
-  H^\prime ( \partial \underline \chi ^\prime )^2 \Big)
\end{equation}

\bigskip

$$  {\underline S}_{\,PV\,1} = - {\partial \over \partial\theta}
 \int d^2x 
   \underline\chi^{\,\ast\,}
{ \tilde {\underline \chi}}  $$

\noindent and the mass term:

\begin{equation} {\underline S}_M = - {1\over 2\pi} M^2 \int d^2x
\underline\chi^{\prime\,^2}
e^{\alpha \,\Omega^\prime}
\end{equation}

The PV superfields involved in this action are:

\begin{eqnarray}
\label{PVSUP2}
{\underline \chi} (x,\theta ) &=& 
\chi (x) + \theta \,
\,\pi^{[\,\chi\,]\,} (x)
\nonumber\\
{\tilde {\underline \chi}} (x,\theta ) &=& \tilde \chi (x) + \theta 
( \,
\,\pi^{[\,\chi\,]} (x) -   c^\prime (x) 
\partial \chi^\prime (x) )\nonumber\\
{\underline \chi}^{\ast} (x,\theta ) &=& \chi^{\ast} (x) + 
\theta B^{\,[\,\chi\,]\,} (x)
\end{eqnarray}

Defining now the total action as 

\begin{equation}
\underline S_T =\underline S + \underline S_{PV} =
S_T + \theta \delta S_T
\end{equation}

\noindent we have:

\begin{equation}
\delta S_T = {1\over 2\pi}  \int d^2x M^2 \chi^{\prime\, 2} 
\Big( (1-\alpha ) \partial c^\prime - \alpha d^\prime \Big)
e^{ \rho^\prime \alpha}
\end{equation}

At this point we arrive at the standard (non superspace) results.
Actually one needs a set of PV fields $\chi^i$ in order to regularize
the above expression. They are, however, all of the same form and, in our 
superspace formulation, will all have actions like (\ref{PVAction}). 
The regularized result (after integrating out the PV fields) is\cite{DJ,JST} :

\begin{eqnarray}
\label{DeltaS} 
& & {1\over \hbar} (i \,\delta S_T\,)_{Reg} = (\Delta S\,)_{Reg}
\nonumber\\ 
&=& {1\over 12\pi}
 \int d^2x 
\Big[ \Big( (1-\alpha ) \partial c^\prime  
- \alpha d^\prime \Big)  
 \Big( \partial^2 h^\prime  
- \alpha (\partial \overline\partial \rho^\prime  ) -
\partial ( h^\prime \partial \rho^\prime  ) \Big) \Big]
\nonumber\\
\end{eqnarray}

The superfield action at one loop order, using this regularization
will then have the general form:

\begin{equation}
{\underline M}_1 ( \alpha ) = M_1  (\alpha) 
+ \theta \Big( i(\Delta S)_{Reg} (\alpha )  + {\bf A} (\alpha ) \Big)
\end{equation}

\noindent where $ M_1  (\alpha)$ is the Wess Zumino term. It is
possible to choose this term in such a way that the anomaly is always
shifted to the trivial symmetry associated to the ghost $d$.  For the
$\alpha = 0$ case the appropriate choice is:

\begin{eqnarray}
{\underline M}_1  (\alpha = 0) &=&{1\over 12\pi}\{  {1\over 2} 
\partial \Omega^\prime\,\,
\overline\partial  \Omega^\prime 
- {1\over 2} H^\prime ( \partial \Omega^\prime )^2
 + H^\prime  \partial^2  \Omega^\prime  \}
\nonumber\\
& &\nonumber\\
&=&{1\over 12\pi}\{ -{1\over 2}  \rho^\prime
\partial\overline\partial \rho^\prime + {1\over 2} \rho^\prime
\partial  h^\prime
\partial  \rho^\prime +  {1\over 2} 
\rho^\prime \, h^\prime \, 
\partial^2 \rho^\prime  +  h^\prime \partial^2 \rho^\prime \} 
\nonumber\\ 
&+&\theta {1\over 12\pi} \Big( - c^\prime
\partial^3 h^\prime  + d^\prime 
\partial^2  h^\prime  
- d^\prime \partial\overline\partial 
\rho^\prime - h^\prime \partial
 d^\prime \partial \rho^\prime  \Big)
\nonumber \\
& &\nonumber\\
&=& M_1 + \theta (i\Delta S + d \overline A) 
\end{eqnarray}

\noindent with 

$$\overline A = {1\over 12\pi}\Big( \partial^2 h^\prime
 - \partial\overline\partial \rho^\prime  +\partial h^\prime 
 \partial \rho^\prime  +
 h^\prime  \partial^2  \rho^\prime \Big)$$

Considering now  $\alpha = 1$, it can be seen from (\ref{DeltaS}) that
in this case the regularization procedure itself leads 
to an anomaly only in the trivial symmetry associated to $d$. We
can thus choose:

\begin{equation}
{\underline M}_1 (\alpha = 1) = 0
\end{equation}

and the anomaly will be:

\begin{eqnarray}
{\bf A} (\alpha =  1 ) &=& - i (\Delta S)_{Reg} ( \alpha = 1 ) \nonumber\\
 &=& {1\over 12\pi} \int d^2x d^\prime  \Big( \partial^2 h^\prime 
-\partial \overline\partial \rho^\prime +  h^\prime \partial \rho^\prime \Big)
\end{eqnarray}

Integration over the auxiliary fields leads to the usual non
superspace results, but it is important to stress that the superspace
formulation requires the presence of the collective fields and the
associated gauge fixing structure.

\section{Conclusion}

Although BRST  superspace formulations for gauge theories have been
known for a while\cite{SUP}, anomalous gauge theories have not yet
been considered in this context.  The Batalin Vilkovisky procedure
represents a very powerful framework for the quantization of this
kind of theories.  We have shown that the (formal) master equation of
the BV formalism can be represented as the requirement of a (formal)
superspace structure for the quantum action.  At one loop order,
using the collective field approach to BV, we have shown that the
Pauli Villars regularization procedure can be translated to
superspace and that the superfield associated to the one loop order
term of the action involves the anomalies and Wess Zumino terms.  An
interesting point that remains as an object of future investigation
is the extension of this superspace formulation for the more general
cases in which anomalies and Wess Zumino terms depend on the
antifields.

\vskip 1cm
\section{Acknowledgements}
The authors  would like to thank Ashok Das for very important
discussions. This work is supported in part by CNPq, FINEP, FAPERJ,
FUJB and CAPES
(Brazilian Research Agencies).
\vfill\eject

\end{document}